\documentstyle[pre,aps,twocolumn,epsf]{revtex}
\def\B.#1{{\bbox{#1}}}
\begin{document}
\title{Convergent Calculation of the Asymptotic Dimension of Diffusion Limited Aggregates:
Scaling and Renormalization of Small Clusters}
\author{Benny Davidovitch, Anders Levermann and Itamar Procaccia}
\address{Dept. of Chemical Physics, The Weizmann Institute of
Science, Rehovot 76100, Israel}

\date{version of \today}
\maketitle
%%%%%%%%%%%%%%%%%%%%%%
\begin{abstract}
Diffusion Limited Aggregation (DLA) is a model of fractal growth 
that had attained a paradigmatic status 
due to its simplicity and its underlying role for a variety of pattern
forming processes. 
We present a convergent calculation of the fractal dimension
D of DLA based on a renormalization scheme for the first Laurent 
coefficient of the conformal map from the unit circle
to the expanding boundary of the fractal cluster.  The theory is applicable from
very small (2-3 particles) to asymptotically large ($n\to \infty$) clusters.  
The computed dimension is $D=1.713\pm 0.003$.
\end{abstract}
\vskip 0.5 cm
The model of Diffusion Limited Aggregation was presented in 1981 by
Witten and Sander \cite{81WS} as a computer algorithm.
The process begins with fixing one particle at the center of 
coordinates in $d$-dimensions, and
follows the creation of a cluster by releasing random walkers from infinity,
allowing them to walk around until they hit any particle belonging to the
cluster. The growing cluster appears to be a random fractal. The model has
attracted enormous attention as an example for the spontaneous creation
of fractal objects in nature, and also as a paradigm for a family of
related ``harmonic" problems that have to do with the solution of the
Laplace equation $\nabla^2 p=0$ with appropriate boundary conditions on
moving, ramified boundaries. Among
such problems are dielectric breakdown \cite{84NPW}, flows in porous media ( with D'arcy's law
$\B.v=-\B. \nabla p$ and $\B.\nabla \cdot\B.v=0$) \cite{85NDS}, electrochemical
deposition \cite{89BMT} etc. 

Numerical estimates 
of the fractal dimension $D$ of DLA \cite{83Mea} 
turned out to converge extremely
slowly with the number of particles $n$ of the cluster, leading even 
to speculations \cite{Mandel} that
asymptotically the clusters were plane filling (i.e. $D=2$ in two dimensions).  
To date there are still
no controlled calculations of the fractal dimension of DLA. The aim of this  
Report is to close this
gap for DLA in two dimensions. We propose a renormalization procedure that culminates 
with an integral equation whose
solution determines the dimension $D$ of DLA. 

To derive the wanted integral equation we use the conformal theory that was
developed recently \cite{98HL,99DHOPSS,99DP,00DFHP,00DP} for fractal growth patterns. In this
theory one considers the conformal map from
the exterior of the  unit circle in the complex plane to the exterior of the growing
fractal cluster. At the basis of this approach lies the understanding that
once a fractal object is well developed, it is extremely difficult to find
a conformal map from a smooth region to its boundary, simply because the conformal
map is terribly singular on the tips of a fractal shape. The derivative of the inverse map is the
growth probability for a random walker to hit the interface (known as
the ``harmonic measure") which has been shown to be a multifractal measure \cite{86HMP}
characterized by infinitely many exponents \cite{83HP,86HJKPS}. Accordingly, in the present
approach one grows the cluster by iteratively constructing the conformal map starting from
a smooth initial interface. Consider 
$\Phi^{(n)}(w)$ which conformally maps the exterior of the unit circle $e^{i\theta}$ in the
mathematical $w$--plane onto the complement of the (simply-connected)
cluster of $n$ particles in the physical $z$--plane \cite{98HL,99DHOPSS,99DP,00DFHP,00DP}. 
The unit circle is
mapped onto the boundary of the cluster. The map $\Phi^{(n)}(w)$ is
made from compositions of elementary maps $\phi_{\lambda,\theta}$,
\begin{equation}
\Phi^{(n)}(w) = \Phi^{(n-1)}(\phi_{\lambda_{n},\theta_{n}}(w)) \ ,
\label{recurs}
\end{equation}
where the elementary map $\phi_{\lambda,\theta}$ transforms the unit 
circle to a circle with a ``bump" of linear size $\sqrt{\lambda}$ around
the point $w=e^{i\theta}$. An example of a good elementary map $\phi_{\lambda,\theta}$ was
proposed in \cite{98HL}, endowed with a parameter
$a$ in the range $0< a < 1$, determining the shape of the bump. We employ $a=1/2$ which is consistent
with semicircular bumps. Accordingly the map $\Phi^{(n)}(w)$ adds on a
new bump to the image of the unit circle under $\Phi^{(n-1)}(w)$. The
bumps in the $z$-plane simulate the accreted particles in
the physical space formulation of the growth process. Since we want
to have {\em fixed size} bumps in the physical space, say
of fixed area $\lambda_0$, we choose in the $n$th step
\begin{equation}
\lambda_{n} = \frac{\lambda_0}{|{\Phi^{(n-1)}}' (e^{i \theta_n})|^2} \ .
\label{lambdan}
\end{equation}
The recursive dynamics can be represented as iterations 
of the map $\phi_{\lambda_{n},\theta_{n}}(w)$,
\begin{equation}
\Phi^{(n)}(w) =
\phi_{\lambda_1,\theta_{1}}\circ\phi_{\lambda_2,\theta_{2}}\circ\dots\circ
\phi_{\lambda_n,\theta_{n}}(\omega)\ . \label{comp}
\end{equation}
The difference between various growth models will manifest itself in the 
different itineraries $\{\theta_1\cdots \theta_n\}$ \cite{00DFHP}.
To grow a DLA we have to choose random positions $\theta_n$.
This way we accrete fixed size bumps in the physical plane according
to the harmonic measure (which is transformed into a uniform
measure by the analytic inverse of $\Phi^{(n)}$).
The DLA cluster is fully determined by the stochastic itinerary
$\{\theta_k\}_{k=1}^n$. In Fig.~\ref{DLAcluster} we present a typical DLA cluster
grown by this method to size $n=$100 000. 

This method affords us analytic power that is lacking
in the original computer algorithm \cite{81WS}; 
The conformal map $\Phi^{(n)}(w)$
is represented in terms of its Laurent expansion:
\begin{equation}
\Phi^{(n)}(w) = F^{(n)}_{1} w + F^{(n)}_{0} + F^{(n)}_{-1}w^{-1} +
F^{(n)}_{-2}w^{-2} + \dots 
\label{eq-laurent-f}
\end{equation}

The recursion equations for the Laurent coefficients of $\Phi^{(n)}(w)$
can be obtained analytically, and in particular one shows that \cite{98HL}
\begin{equation}
F_1^{(n)}  = \prod_{k=1}^{n}  \sqrt{1+\lambda_k}  \ , \quad({\rm with ~the ~choice}~ a=1/2 )\ .
\label{F1a}
\end{equation}
The first Laurent coefficient $F_1^{(n)}$ has a distinguished role in determining
the fractal dimension of the cluster, being identical to the
Laplace radius which is the radius of a charged disk having the
same field far away as the charged cluster \cite{99DHOPSS}. Moreover, 
defining $R_n$ as the minimal radius of 
all circles in $z$ that contain the $n$-cluster, one can prove that \cite{83Dur}
\begin{equation}
R_n\le 4F_1^{(n)} \ . \label{RvsF}
\end{equation}
Of course, for every realization $\{\theta_i\}_{i=1}^n$ the first
Laurent coefficient is a random number depending on $\{\theta_i\}_{i=1}^n$ and
on $\lambda_0$. It is thus natural to consider
the mean of $F_1^{(n)}$ over all the possible realization of growth:
\begin{equation}
\langle F_1^{(n)}\rangle(\lambda_0) \equiv \int_0^{2\pi} d\theta_1\cdots \int_0^{2\pi} d\theta_n
F_1^{(n)}(\{\theta_i\}_{i=1}^n,\lambda_0)
\end{equation}

In light of Eq.(\ref{RvsF}) one expects that for
sufficiently large clusters
\begin{equation}
\langle F^{(n)}_1\rangle(\lambda_0) \sim \sqrt{\lambda_0}~n^{1/D} \ , \quad n\to \infty \ .
\label{eq-scalingrad}
\end{equation}
But this is true only for very large values of $n$. For arbitrary values of
$n$ we offer the following proposition, which is central to our developments:
\vskip 0.5 mm {\bf proposition} : For $\lambda_0$ of $O(1)$
$\langle F^{(n)}_1\rangle(\lambda_0)$ is a scaling function of the single
variable $x\equiv \sqrt{\lambda_0} ~(n+\alpha)^{1/D}$, where $n$ takes
on integer values $n=1,2,3\cdots$
and $\alpha$ is a weak function of $\lambda_0$, taking on values of the
order of unity.
\vskip 0.5 mm
The origin of the parameter $\alpha$ is evident - it stems from the fact that
for varying values of $\lambda_0$ the unit circle around which we grow
the bumps is contributing to the Laplace radius, changing the effective value
of $n$. We will show however that to very good approximation
$\alpha$ can be taken as constant.

We demonstrate the proposition by a direct calculation of $\langle F^{(n)}_1\rangle(\lambda_0)$.
In Fig.2 panel a we show $\langle F^{(n)}_1\rangle(\lambda_0)$ for ten different
values of $\lambda_0$ as a function of $n$. In panel b we show the same data as
a function of the scaling variable $x$ using $\alpha=0.2$, $D=1.71$. The data collapse
is evident. 

We stress three points: (i) The scaling function appears to exist for
all values of $n$ starting from $n=1$. It is a {\em non-linear} function of
the scaling variable, with the attainment of the linear regime (\ref{eq-scalingrad}) not
in sight. (ii) $\alpha$ was taken as a
$\lambda_0$-independent constant; Even better data collapse could be obtained with a 
$\lambda_0$-dependent $\alpha$, and we return to this issue below. (iii) The attainment of data
collapse requires a value for $D$; we have used $D=1.71$ but close-by values would have done equally
well by  changing $\alpha$ a bit. Thus we cannot propose the data collapse as an accurate method
of computing the fractal dimension $D$. For this purpose we derive now an
integral equation from which both $\alpha$ and $D$ can be computed in a controlled
fashion.

Having a scaling function in mind we think about a 
real-space renormalization group
procedure, in which we change the number of particles in the cluster
and their size such as to keep the cluster invariant. For $n$ very large,
when the radius of gyration is linear in $\sqrt{\lambda_0}~n^{1/D}$, we can change  
the size of the particles $\sqrt{\lambda_0}$ by a factor of 2, and their number
$n$ by a factor of $2^{D}$. In the asymptotic regime such a renormalization
will leave $R_n$ invariant. For small values of $n$ we make use of the discovery that 
$\langle F_1^{(n)}\rangle(\lambda_0)$ is a scaling function of the single
variable $x=\sqrt{\lambda_0}(n+\alpha)^{1/D}$. We demand that upon renormalization
the average Laplace radius remains invariant. In other words, the fixed point
condition can be written as the following integral equation:
\begin{eqnarray}
&& \int_0^{2\pi} d\theta_1\cdots \int_0^{2\pi} d\theta_n
F_1^{(n)}(\lambda_0) \nonumber\\&&=  \int_0^{2\pi} d\theta_1\cdots \int_0^{2\pi} d\theta_{\bar n}
F_1^{(\bar n)}(\bar \lambda_0) \ , \label{inteq}
\end{eqnarray}
where for any $\bar n> n$ the equation is satisfied by the
unique value of $\bar\lambda_0$ that solves the equation
\begin{equation}
\sqrt{\bar \lambda_0} = \sqrt{\lambda_0} \left(\frac{n+\alpha}{\bar n+\alpha}\right)^{1/D} \ .
\label{lambar}
\end{equation}
The way to compute $D$ is then obvious - one computes the integral on the LHS
of Eq.(\ref{inteq}) for some value of $n$, and then finds the unique value
of $\bar \lambda_0$ for which the RHS with $\bar n> n$ equals the LHS. Then
\begin{equation}
D=2\frac{\ln(\bar n+\alpha)-\ln(n+\alpha)}{\ln{\lambda_0}-\ln{\bar\lambda_0}} \ . \label{D}
\end{equation}
One should stress that Eq.(\ref{inteq}) is all explicit - there is no simulation
or randomness left at this point - the integrand on both sides is an explicit
function of $\{\theta_1,\cdots,\theta_n\}$ or $\{\theta_1,\cdots,\theta_{\bar n}\}$
through Eqs.(\ref{F1a}) and (\ref{lambdan}). We note that this method of calculation
is fundamentally different from the standard method of log-log plots of the radius
of the cluster vs. $n$ \cite{83Mea,Mandel}. These rely on the proportionality of
$R_n$ and $n^{1/D}$. Here we may use {\em small values of} $n$ as we depend 
neither on the asymptotic linearity of (\ref{eq-scalingrad}), nor on self-averaging.
 
Clearly the integral equation will be useful for
an actual determination of $D$ only if it converges to the fixed point {\em quickly} upon
increasing $n$. Otherwise the calculation of the multidimensional integral will become very cumbersome,
maybe even more time consuming than standard numerical simulations. In the rest
of this Report we demonstrate that the calculation actually 
converges very quickly and present the determination of $D$.

Firstly we determine the value of $\alpha$. We use $\bar\lambda_0=1$, and 
solve Eq.(\ref{inteq}) for the simplest case $n=1$, $\bar n=2$, demanding
\begin{eqnarray}
\FL
\langle F_1^{(2)}\rangle(\bar\lambda_0=1)&=&\sqrt{2}\langle\sqrt{1+\bar\lambda_2}\rangle
=F_1^{(1)}(\lambda_0)= \sqrt{1+\lambda_0}\\
(2+\alpha)^{1/D}&=&(1+\alpha)^{1/D}\sqrt{\lambda_0}
\end{eqnarray}
Computing explicitly (taking $D=1.71$) we found $\langle\sqrt{1+\bar\lambda_2}
\rangle\approx 1.241$ from
which followed $\alpha\approx 0.146$. To bracket the calculations we have 
employed this value together with $\alpha=0$ and $\alpha=1$.

In solving Eq.(\ref{inteq}) we chose invariably $\bar n=n+1$, and $\lambda_0=1$.
The results are shown in Fig.3, in the form of the value of $D$ vs. $(n+\alpha)^{-1}$.  Fitting
the best  nonlinear curve to the data with $\alpha=0.146$  we find that it extrapolates
for $(n+\alpha)^{-1}\to 0$ to $D=1.7150...$. Obviously the calculation with $\alpha=0$ and $\alpha=1$
bracket this  from above and from below. Nevertheless the best nonlinear fits to these
data extrapolate to very
close values for $(n+\alpha)^{-1}\to 0$ (see Fig.3). Taking all the data together we can
present a final value of $D=1.713\pm 0.003$. We cannot overstress the fact that
these results were obtained from solving Eq.(\ref{inteq}) with $n$ values ranging
from $n=1$ to $n=20$. We believe that this represents a major advance compared
to traditional estimates of the dimension of DLA.

To improve the results even further, and to remove the curvature in the line
of $D$ vs. $(n+\alpha)^{-1}$ we can endow $\alpha$ with a weak $\lambda_0$
dependence. Using the function
\begin{equation}
\alpha(\lambda_0) =  2.04811  - 4.08071 \sqrt{\lambda_0} + 2.23446 \lambda_0  \ , \label{alpha}
\end{equation}
and using the appropriate value of $\alpha(\lambda_0)$ in Eq.(\ref{D}) we
obtain solutions with essentialy constant $D$ for all values of $n>2$.
The values of $D$ computed are shown in Fig.4 as a function of $n$. With this
data we can state without going to the limit $n\to \infty$ that $D=1.713\pm 0.005$. This result
is invariant with respect to changing $n$ in Eq.(\ref{inteq}). It can be
computed equally well from $n=3$ and $\bar n=4$ as from  and $n=19$ and $\bar n=20$.
We should of course stress at this point that the function $\alpha(\lambda_0)$
was {\em not} computed from first principles; we hope that further theoretical
progress will shed light on how to achieve {\em a-priori} determination
of the functional form $\alpha(\lambda_0)$.

The most important questions that we need to address now are: (i)
why the classical numerical estimates \cite{83Mea,Mandel} of the fractal dimension of DLA 
converge so slowly, whereas here we can get an excellent estimate of $D$
even with $n$ of the order of unity. (ii) Is the collapse of the scaling
data for small $n$ exact or an excellent approximation. The answer to (i) is that
in standard numerical experiments the radius of gyration of the grown cluster
was plotted in log-log coordinates against the number of particles, with
$D$ estimated from the slope. Examining our scaling function  $\langle F_1\rangle (x)$ 
(see Fig. 2) we note the slow crossover to linear behavior, which may not be
fully achieved even for extremely high values of $n$. In this respect we understand
that a reliable estimate of $D$ from radius of gyration
calculation requires inhuman effort, as was indeed experienced by workers in the
field \cite{Mandel}. In the present formulation the appearance of the {\em asymptotic} $D$
as a renormalization exponent already at early stages of the growth allows
a convergent calculation. The answer to (ii) has to await a first principle theory
for $\alpha(\lambda_0)$.

\acknowledgments
It is a pleasure to acknowledge useful discussions with Ayse Erzan, Mitchell  Feigenbaum
and George  Hentschel. 
This work has been supported in
part by the European Commission under the TMR program, The Petroleum Research Fund,
The Minerva Foundation, Heidelberg, Germany
and the Naftali and Anna 
Backenroth-Bronicki Fund for Research in Chaos and Complexity.
%%%%%%%%%%%%%%%%%%%%%%

%%%%%%%%%%%%%%%%%%%%%%%%%%%%%%%%%%%%%%%%%%%%%%%%%%%%%%%%%%%%%%%%%%%%
%\narrowtext
%\begin{figure}
%\hskip -1.5 cm
%\epsfxsize=6.0truecm
%\epsfysize=4.0truecm
%\epsfbox{deterDLA100000.ps}
%\caption{A DLA cluster, $n=100 000$.}
%\label{DLAcluster}
%\end{figure}
%%%%%%%%%%%%%%%%%%%%%%%%%%%%%%%%%%%%%%%%%%%%%%%%%%%%%%%%%%%%%%%%%%%
%%%%%%%%%%%%%%%%%%%%%%%%%%%%%%%%%
%%%%%%%%%%%%%%%%%%%%%%%%%%%%%%%%%%%%%%%%%%%%%%%%%%%%%%%%%%%%%%%%%%%%
\narrowtext
\begin{figure}
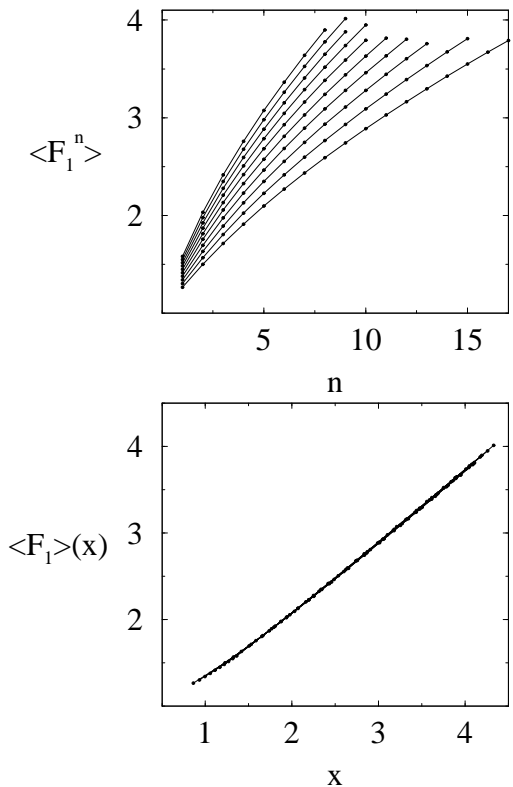

\epsfxsize=7.5truecm
\epsfbox{DLAanalyticFig.2a}
\vskip -1.0 cm
\epsfxsize=7.5truecm
\epsfbox{DLAanalyticFig.2b}
\caption{Panel a: the mean $\langle F_1^{(n)}\rangle$ as a function of $n$ for 
different values of $\lambda_0$. The data shown pertain to the 10 values of $\lambda_0$
from 0.6 (below) to 1.5 (above). Panel b: the
same objects plotted as a function of the scaling variable $x$ with $\alpha=0.2$
and $D=1.71$.} 
\label{meanF}
\end{figure}
%%%%%%%%%%%%%%%%%%%%%%%%%%%%%%%%%%%%%%%%%%%%%%%%%%%%%%%%%%%%%%%%%%%
%%%%%%%%%%%%%%%%%%%%%%%%%%%%%%%%%%%%%%%%%%%%%%%%%%%%%%%%%%%%%%%%%%%%
\narrowtext
\begin{figure}
\hskip 0.2 cm
\epsfxsize=8.0truecm
\epsfbox{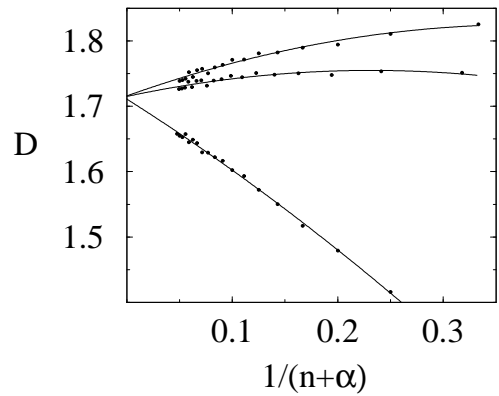}
\caption{The estimated value of the fractal dimension $D$ computed from 
the solution of the integral equation (9). The three sets of points pertain
to $\alpha=0$ (upper), $\alpha=0.146$ (middle) and $\alpha=1$ (lower). The lines are the best
quadratic fits to the data. The extrapolated values at $1/(n+\alpha)=0$ are
1.7158.., 1.7150.. and 1.7111.. respectively, leading to our final estimate $D=1.713\pm 0.003$}
\label{Dcalculation}
\end{figure}
%%%%%%%%%%%%%%%%%%%%%%%%%%%%%%%%%%%%%%%%%%%%%%%%%%%%%%%%%%%%%%%%%%%
%%%%%%%%%%%%%%%%%%%%%%%%%%%%%%%%%%%%%%%%%%%%%%%%%%%%%%%%%%%%%%%%%%%%
\narrowtext
\begin{figure}
\hskip 0.2 cm
\epsfxsize=8.0truecm
\epsfbox{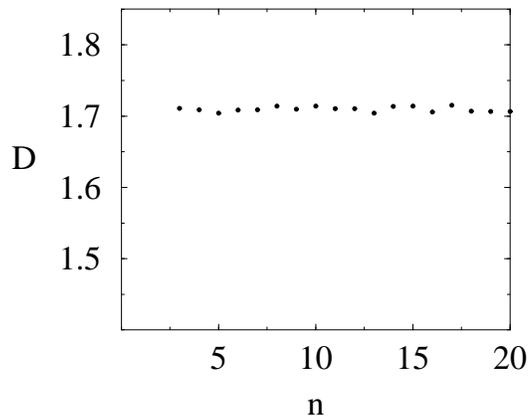}
\caption{The estimated value of the fractal dimension $D$ computed from 
the solution of the integral equation (9), using the function $\alpha(\lambda_0)$
of Eq.(14). Without going to the limit $n\to \infty$ we
can state $D=1.713\pm 0.005$}
\label{Dcalcalpha}
\end{figure}
%%%%%%%%%%%%%%%%%%%%%%%%%%%%%%%%%%%%%%%%%%%%%%%%%%%%%%%%%%%%%%%%%%%

%\end{multicols}

\end{document}